\documentclass[epj]{svjour}
\usepackage{graphics}
\usepackage[pdftex]{graphicx}
\usepackage{xspace,colortbl}
\usepackage{amsmath}
\usepackage{amssymb}
\usepackage[T1]{fontenc}
\usepackage{float}
\usepackage[utf8]{inputenc}
\usepackage{authblk}
\usepackage{graphicx}

\begin{document}
\title{Accretion disk around regular black holes}

\author{Amin Rezaei Akbarieh\inst{1} \thanks{am.rezaei@tabrizu.ac.ir} \and Minou Khoshragbaf\inst{1}
\thanks{mo.khoshrang@tabrizu.ac.ir} \and
Mohammad Atazadeh\inst{2}
\thanks{atazadeh@azaruniv.ac.ir}%
}                     
%
%
\institute{ Faculty of Physics, University of Tabriz, Tabriz 51666-16471, Iran \and Department of Physics, Azarbaijan Shahid Madani University, Tabriz, 53714-161 Iran}
\date{Received: date / Revised version: date}

\abstract{
Regular black holes are crucially important as approaches to solving the singularity problem, and in this paper, the accretion disk of Bardeen and Hayward models have been studied. For this purpose, we calculated the physical properties of black holes, including radiant energy, luminosity derivative, temperature, and conversion efficiency of accretion mass into radiation. The obtained results show that the non-zero-free parameters of regular black holes cause the radius of the innermost stable circular orbit of the disk to shift to smaller values. As a result of this displacement, we saw an increase in the profiles of radiant energy, luminosity derivative, and temperature. We also find that Bardeen and Hayward's black holes are more efficient in converting mass to radiation than Schwarzschild. Finally, we compared the free parameter of these two black holes with the spin of the rotating black hole and found that the Bardeen and Hayward black holes can mimic the slowly rotating Kerr black hole.
\PACS{
      {PACS-key}{discribing text of that key}   \and
      {PACS-key}{discribing text of that key}
     } 
} 
\maketitle
\section{Introduction}\label{intro}
General relativity can make predictions about some astrophysical objects \cite{Berti:2015itd}, such as black holes \cite{BarackLeor:2019mnu}. They are considered very interesting in astrophysics and have been studied in a broad range of masses \cite{EventHorizonTelescope:2019ths}. Due to their causal structure, black holes have a surface to which any particle or wave that passes through this surface cannot return; this is known as the event horizon \cite{Rodrigues:2022qdp}. Black holes are regions of spacetime that have a singularity problem and are gravitationally collapsed \cite{Frolov:2013efa}. Since the inner regions of the event horizon of black holes have singularities, a classical concept of spacetime can no longer be defined. This is why general relativity is considered a theory only valid for certain energy scales. After more than a century from Schwarzschild, obtaining a comprehensive description of black holes remains at the heart of fundamental questions in the unification of general relativity and quantum mechanics \cite{Penrose:1964wq,Hawking:1976ra}.\\
To solve the singularity problem, one could remove the singularities from general relativity in proper astrophysical regions via different methods and extract relevant astrophysical observables. Due to technological advancement, observational and experimental methods could play a critical role in verifying theoretical predictions about astrophysical objects \cite{SimpsonAlex:2022mnu}. The direct observations of gravitational waves from an astrophysical source in LIGO/Virgo merger events \cite{LigoCaltechEdu:2022mnu,ListofGravitationalWaveObservations:2021mnu}, the Event Horizon Telescope (EHT) image of the black hole at the center of the galaxy M87 \cite{EventHorizonTelescope:2019ths,EventHorizonTelescope:2019pcy,EventHorizonTelescope:2019uob,EventHorizonTelescope:2019jan,EventHorizonTelescope:2019pgp,EventHorizonTelescope:2019ggy} and the black hole Sgr A* at the center of the Milky Way \cite{EventHorizonTelescope:2022vjs,EventHorizonTelescope:2022xqj,EventHorizonTelescope:2022xnr,EventHorizonTelescope:2022exc,EventHorizonTelescope:2022urf,EventHorizonTelescope:2022gsd,EventHorizonTelescope:2022tzy,EventHorizonTelescope:2022okn,EventHorizonTelescope:2022ago,EventHorizonTelescope:2022wok} are some of the remarkable outcomes of the significant advances in technology. Some other studies conducted in different fields related to observable astrophysical quantities can be found in \cite{Eiroa:2012fb,FlachiAntonino:2013mnu,Abdujabbarov:2016hnw,Carballo-Rubio:2018pmi,Carballo-Rubio:2019nel,Carballo-Rubio:2020ttr,Dai:2019nph,Cramer:1994qj,Simonetti:2020ivl,Berry:2020ntz,Carballo-Rubio:2021ayp,Carballo-Rubio:2021wjq,Bronnikov:2021liv,Churilova:2021tgn,Bambi:2021qfo,Simpson:2021biv}. Therefore, experiments and observations guide theoretical physicists to correct, adjust, or discard their theories, and this is the step-by-step approach to achieving complete theories.\\
One way to solve the problem of the existence of singularities is to replace the interior of the single black hole with a singularity-free core or use so-called regular black holes \cite{Simpson:2021dyo,Berej:2006cc}. These spacetimes have event horizons, but at the same time, they lack any pathological features such as singularities or regions with closed time curves \cite{Bambi:2013ufa}. Notice that regular black holes are not vacuum solutions of Einstein's gravitational equations, but they necessarily contain an additional field or satisfy a form of modified gravity theory. Therefore, they violate the energy conditions associated with the existence of physical singularities \cite{Stuchlik:2014qja,Hawking:1973uf}. The idea of regular black holes emerged in the mid-60s and is still popular today \cite{BardeenJ:1968mnu,Bardeen:1972fi,Sakharov:1966aja}. Since James Bardeen first proposed regular black holes, they are also known as Bardeen black holes \cite{BardeenJ:1968mnu}. The Bardeen model is a regular spacetime black hole that satisfies the weak energy condition \cite{Ayon-Beato:2000mjt}, and this condition is valid for all regular black holes \cite{Borde:1996df}. Because of this weak energy condition, regular black holes avoid singularity theorems. Unlike other black holes, the core of regular black holes does not have a singularity; however, the environment outside the event horizon in regular black holes is similar to other black holes \cite{Berej:2006cc}. What Bardeen means by regular is regularity, which is obtained by applying a global constraint on the components of the conventional curvature tensor and Riemann curvature variables; in other words, the black hole formula has no discontinuity \cite{Simpson:2021dyo}. After Bardeen, other regular black hole models \cite{Borde:1994ai,Barrabes:1995nk,Mars:1996khm,Hayward:2005gi,Cabo:1997rm} have been investigated regarding spherical and axial symmetry \cite{Abdujabbarov:2016hnw,Carballo-Rubio:2018pmi,Carballo-Rubio:2019nel,Carballo-Rubio:2020ttr,Carballo-Rubio:2021ayp,Carballo-Rubio:2021wjq,Bambi:2013ufa,BardeenJ:1968mnu,Ayon-Beato:2000mjt,Hayward:2005gi,Bronnikov:2006fu,Li:2013jra,Amir:2018pcu,Soroushfar:2021mis,Jusufi:2020odz,Herdeiro:2016tmi,Frolov:2014jva,Amir:2016cen,Neves:2014aba,Toshmatov:2014nya,Tinchev:2015apf,Fan:2016hvf,Toshmatov:2017zpr,Toshmatov:2017anu,Simpson:2019mnu,Simpson:2019cer,Lobo:2020ffi,Simpson:2019mud,Brahma:2020eos,Mazza:2021rgq,Franzin:2021vnj}.\\
We are lucky that under normal astrophysical conditions, a black hole of any size and mass is rarely naked, and in most cases, the black hole is covered with gaseous material. This gas, drawn in a spiral motion, forms a hot accretion disk that emits a distinct spectrum of electromagnetic radiation \cite{Churilova:2021tgn}. The radiation of black holes is due to accretion disks around them. The accretion disk of the black hole is spiraled towards the central black hole by the action of viscosity; it sheds its initial angular momentum outwards and releases the gravitational potential energy in the form of heat \cite{Liu:2022cph}. Some or all of the released heat is radiated, which produces diverse spectra depending on the specific radiative processes. Four basic models to describe accretion disks are: 1) the standard thin disk, 2) the optically thin two-temperature disk, 3) the slim disk, and 4) the advection-dominated accretion flow\cite{Shakura:1972te,SHAPIRO:1989mnu,KatzJ:1977mnu,Begelman:1978mnu,Abramowicz:1988mnu,Ichimaru:1977mnu,Rees:1982pe,Narayan:1994is,Narayan:1994xi,Narayan:1994et,Chen:1995uc,Abramowicz:1996ww}. Note that a combination of these methods can be used for describing black holes accretion disks.\\
Due to the critical importance of black holes as a test site for theories of gravity in strong field regions, researchers have a great incentive to study the physics of black holes. At the same time, we can observe the immediate environment of black holes with high resolution and comparable to the event horizon through the event horizon telescope. In this way, it is possible to fill the gap between theory and observations by studying the recorded images of black hole accretion and examining the shape and size of the shadow of rotating and non-rotating black holes \cite{Held:2019xde,Lu:2019ush,Kumar:2019ohr,Eichhorn:2021iwq}.\\
Therefore, this motivates us to investigate the accretion disk around Bardeen and Hayward regular black holes and extract their thermal properties. Also, by comparing the extracted results with the results obtained from the observations, we aim to take an effective step forward in better understanding the mechanism governing black holes and accretion disks. \\
Besides the introduction, this article includes the following sections: 2. Hayward and Bardeen black holes, 3. Relativistic thin accretion disk, 4. Conversion efficiency of mass to radiation, 5. Black hole with non-zero spin, and 6. Conclusion

\section{Hayward and Bardeen black holes}

\begin{figure}
\centering
\includegraphics[width=1\linewidth]{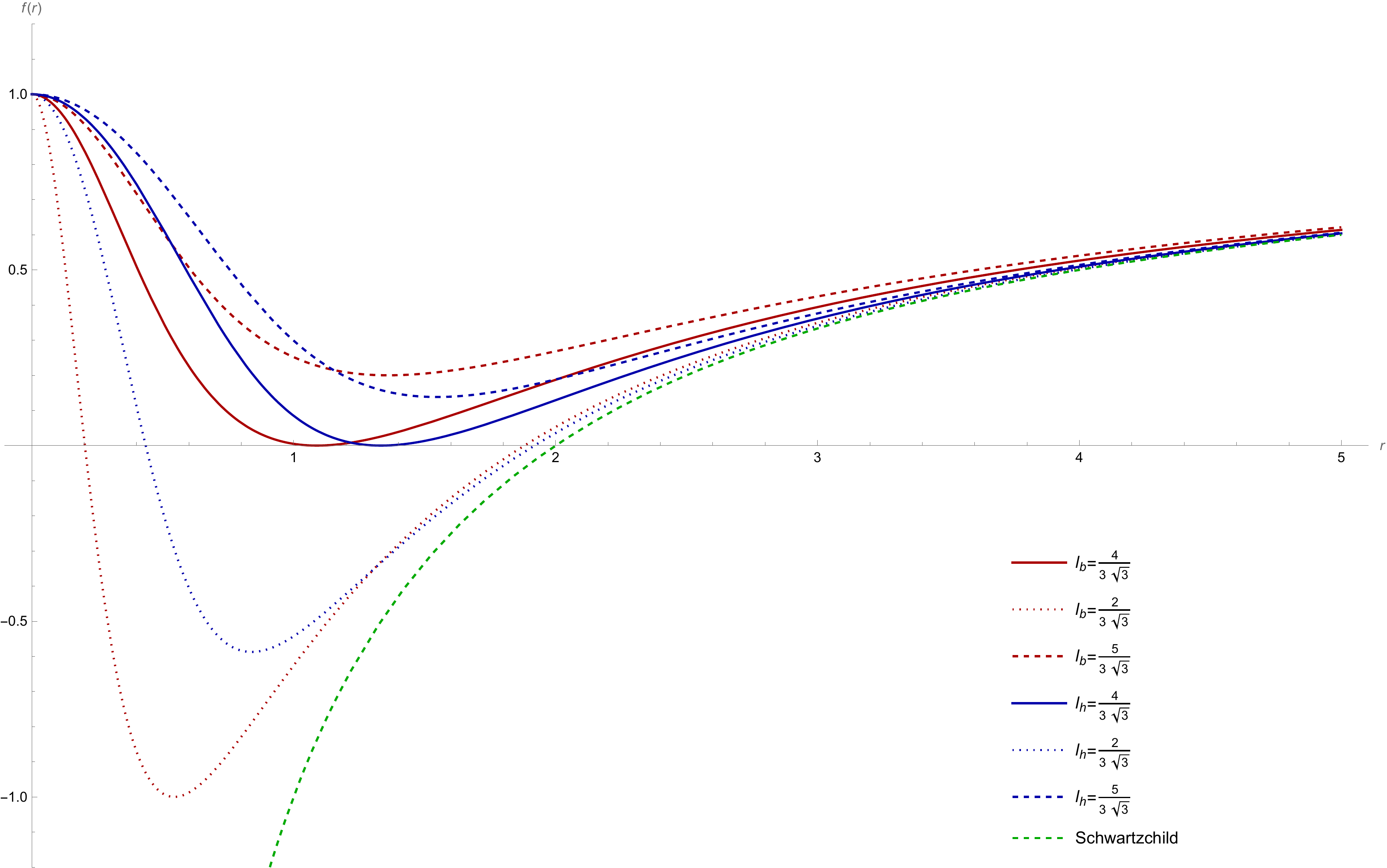}
\caption{Plot of modified metric coefficient $f(r)$ for different values of the free parameter $\ell_b$ for Bardeen (red color) and $\ell_h$ Hayward black holes (blue color). The lines (red and blue) show the modified metric per critical state $\ell_b=\ell_h=4/3\sqrt{3}$. The dashed lines (red and blue) show the modified metric for free parameters greater than the threshold value $\ell_b=\ell_h=5/3\sqrt{3}$. Dashed dots (red and blue) show the modified metric for free parameters lesser than the critical value $\ell_b=\ell_h=2/3\sqrt{3}$. Finally, the green dashed line shows the classical metric $f_0(r)$ for $m=1$.}
\label{fig001}
\end{figure}

Bardeen and Hayward black holes are spherically symmetric, asymptotically flat, static. The Einstein tensor obtained for these black holes is physically reasonable and they have regular centers. In addition, they satisfy the weak energy conditions and have components that are confined over large distances \cite{Hayward:2005gi}. They have no pathological features such as singularities or regions with closed timelike curves and have a horizon \cite{Bambi:2013ufa}.\\
The spherical and static solution of the black hole is generally as follows
\begin{eqnarray}
\label{eq.01}
ds^2=-f(r)dt^2+\frac{dr^2}{f(r)}+r^2(d\theta^2+\sin^2{\theta}d\phi^2),
\end{eqnarray}
where generally $f(r)$ is
\begin{eqnarray}
\label{eq.02}
f(r)=1-\frac{2M_{eff}(r)}{r}.
\end{eqnarray}
In the case $M_{eff}(r)=m$, the Schwarzschild solution will be obtained. Different choices for $f(r)$ can lead to other solutions, for Bardeen and Hayward's black holes in $c=G_0=1$ units, we have
\begin{eqnarray}
\label{eq.03}
    f_b(r)=1-\frac{2m r^3}{(r^2+\ell_b^2)^{3/2}},
\end{eqnarray}
and
\begin{eqnarray}
\label{eq.04}
    f_h(r)=1-\frac{2m r^3}{r^3+2 m \ell_h^2},
\end{eqnarray}
where the indices "b" and "h" stand for Bardeen and Hayward, respectively. Trivially, in the limit $\ell_b , \ell_h\rightarrow 0$, the Schwarzschild space-time is recovered. The radius of the event horizon for each black hole is obtained from $f(r)=0$
\begin{eqnarray}
\label{eq.05}
  (r^2+m^2\bar{\ell}_b^2)^{3/2}-2m r^2=0,
\end{eqnarray}
and
\begin{eqnarray}
\label{eq.06}
    r^3+2m^3\bar{\ell}^2_h-2m r^2=0,
\end{eqnarray}
in which $\bar{\ell}_b=\ell_b/m$ and $\bar{\ell}_h=\ell_h/m$. Considering that the critical value for the existing of the event horizon is $\bar{\ell}_b=\bar{\ell}_h=4/3\sqrt{3}$, for $\bar{\ell}_b,\ bar{\ell}_h>4/3\sqrt{3}$ there is no event horizon and in this case a naked singularity is created. Since the plan of this work is to investigate the radiative properties of accretion discs around regular black holes, we are not dealing with singularities.\\
In Fig. \ref{fig001}, we plotted $f(r)$ in terms of $r$ for both Bardeen and Hayward black holes with different values of the free parameters. The red color is for Bardeen black hole, and the blue color shows the outcomes of Hayward black hole. Furthermore, the continuous lines demonstrate the critical values of the free parameters, and the dotted lines accord to lesser than the critical value. Eventually, the dashed lines show the values greater than the critical value of the free parameter. In order to be able to compare more easily, the classical metric coefficient $f_0(r)=1-2m r^{-1}$ for $m=1$ is shown as a green dashed line. It is obvious the event horizons of regular black holes are shifted to lower values compared to the classic black holes. From Fig. \ref{fig001}, One can conclude that for critical values $\bar{\ell}_b=\bar{\ell}_h=4/3\sqrt{3}$, both black holes have an one event horizon, for larger parameters, there is no horizon and for smaller ones, they have two event horizons. It should be noted that the red and the blue dashed lines correspond Bardeen and Hayward black hole, respectively. Besides, in general, for a given value of $m$ and the same free parameters $\bar{\ell}_b=\bar{\ell}_h$, the horizon of Bardeen black hole (red color) occurs in lesser amounts compared to the Hayward (blue color). In addition, as expected from regular black holes, in the limit $r\rightarrow 0$, the value of both black holes approaches $f(r)=1$ and for $r\rightarrow\infty$, the behavior of Bardeen and Hayward resembles to Schwarzschild black hole.\\

\begin{figure}
\centering
\includegraphics[width=1\linewidth]{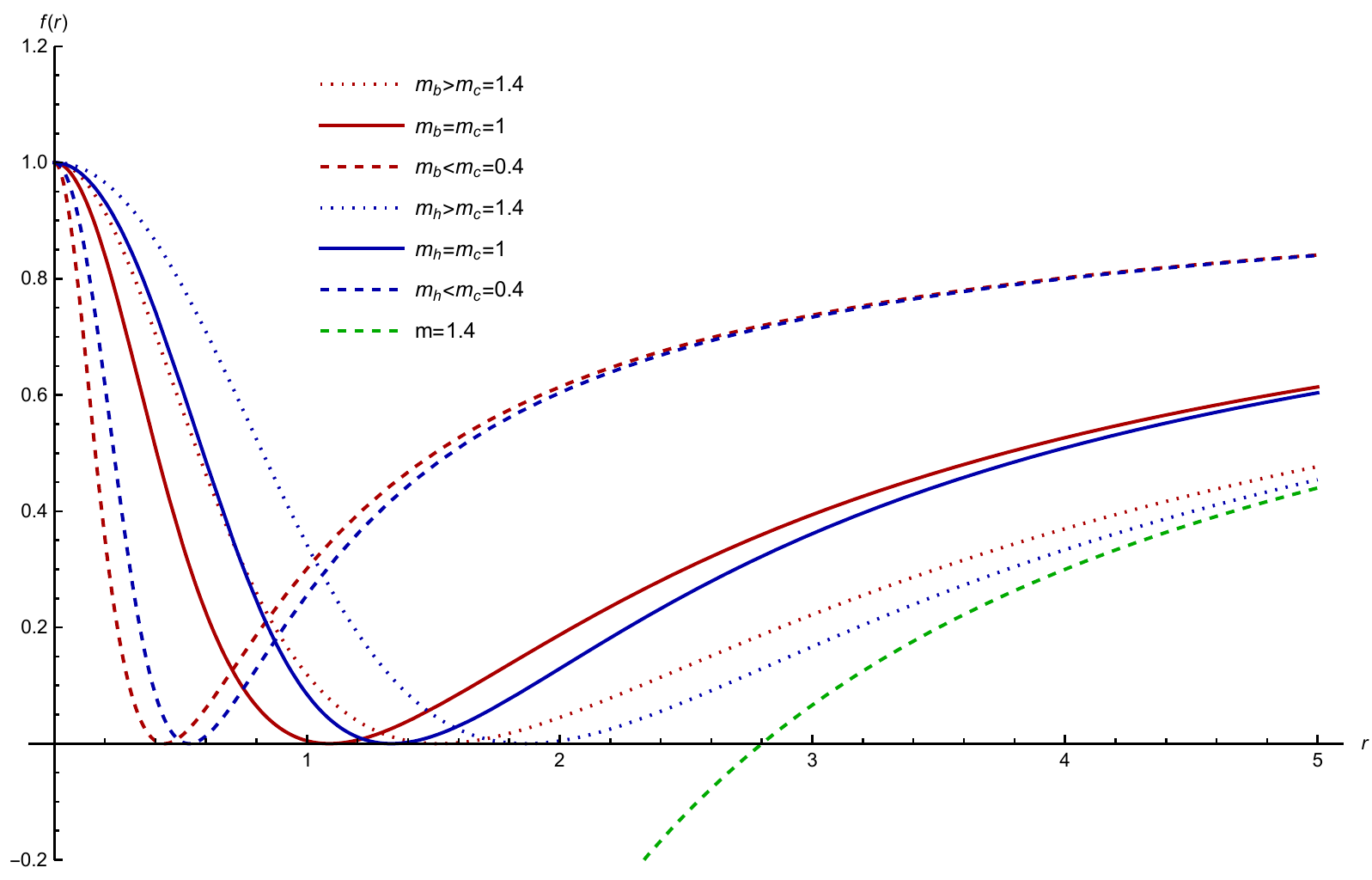}
\caption{Plot of modified metric coefficient $f(r)$ for different values of mass $m$ and constant value of free parameters $\ell_b=\ell_h=4/3\sqrt{3}$ for Bardeen (red color) and Hayward black holes (blue color). The solid lines (red and blue) show the modified metric for the mass limit state $m=m_c$. The dashed lines (red and blue) show the modified metric for masses larger than the mass limit $m>m_c$. The dashed dots (red and blue) show the modified metric for masses smaller than the critical mass $m<m_c$. Finally, the green dashed line shows the classical metric $f_0(r)$ for $m=1.4$.}
\label{fig002}
\end{figure}

\begin{figure}
\centering
\includegraphics[width=1\linewidth]{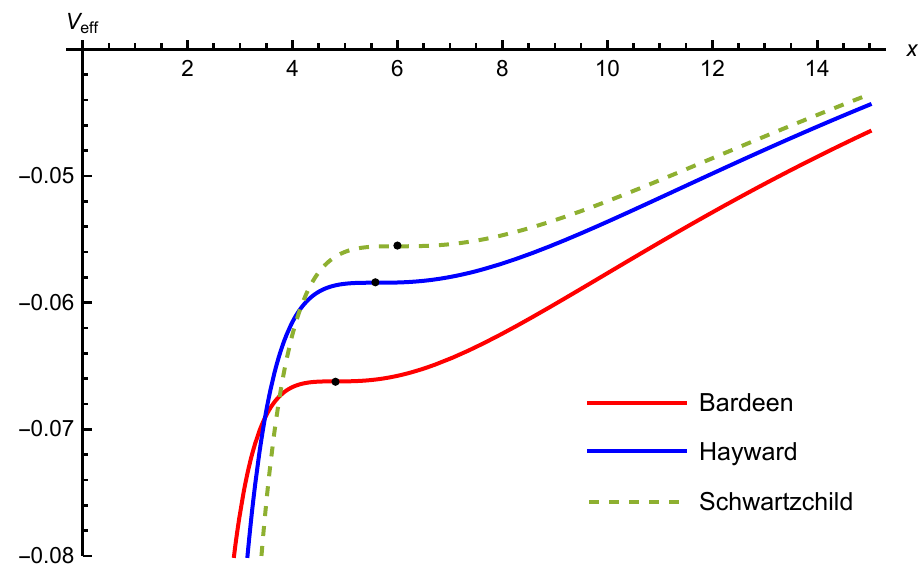}
\caption{The effective potential for the constant value of the angular momentum in the ISCO is drawn for the critical state of free parameters $\bar{\ell}_b=\bar{\ell}_h=\frac{4}{3\sqrt{3}}$ (Hayward blue, Bardeen red) and for the classical state $\bar{\ell}_b=\bar{\ell}_h=0$ (green dashed curve). The black dots indicate the locations of the ISCO for each of the listed cases.}
\label{fig003}
\end{figure}

\begin{figure}
\centering
\includegraphics[width=1\linewidth]{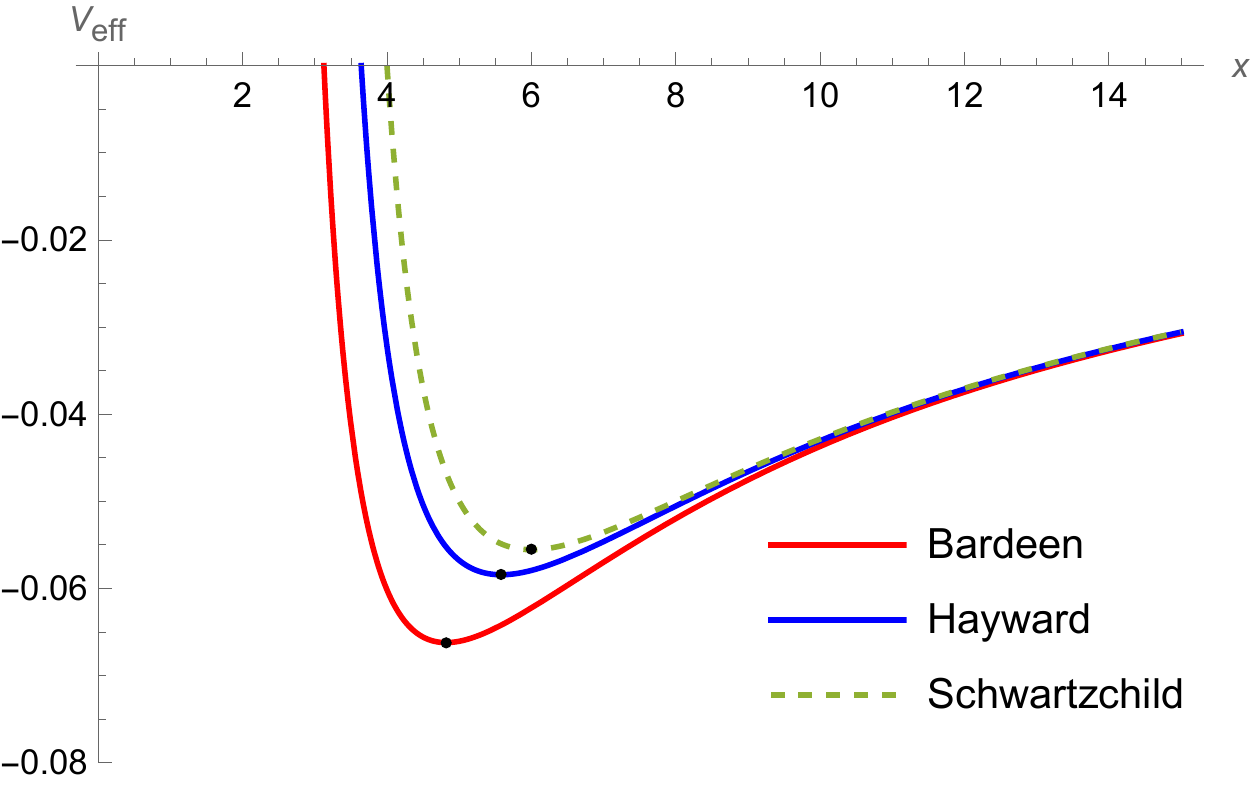}
\caption{Effective potential as a function of $x$ for Bardeen (red) and Hayward (blue) black holes for the free parameters $\bar{\ell}_b=\bar{\ell}_h=\frac{4}{3\sqrt{3}}$. For direct comparison, the classical case $\bar{\ell}_b=\bar{\ell}_h=0$ (dashed curve) is also shown.}
\label{fig004}
\end{figure}

In Fig. \ref{fig002}, $f(r)$ in terms of $r$ is plotted, and since there is no horizon for the values $\ell_b=\ell_h>4/3\sqrt{3}$, we have drawn the function $f(r)$ for $\ell_b=\ell_h=4/3\sqrt{3}$. It is clear that by varying the mass, only the position of the event horizon changes, and the number of them is independent of the mass value. Therefore, for values lesser than the critical mass compared to the critical mass, the event horizon of the black hole shifts towards lesser values of $r$ and conversely. Also, for fixed $m$, the event horizon of a Bardeen black hole is smaller compared to a Hayward black hole.\\
By introducing the "effective" mass $M_{eff}(r)$ which is a function of the radius, the line element in Eqs. (\ref{eq.03}) and (\ref{eq.04}) is
\scriptsize{\begin{eqnarray}
\label{eq.07}
   ds^2 = (1-2\frac{M_{eff}(r)}{r})dt^2+(1-2\frac{M_{eff}(r)}{r})^{-1}dr^2+r^2d\Omega^2,
\end{eqnarray}}\normalsize
where 
\begin{eqnarray}
\label{eq.08}
   M_{eff-b}(r)=\frac{2m r^3}{(r^2+\ell_b^2)^{3/2}},
\end{eqnarray}
and
\begin{eqnarray}
\label{eq.09}
   M_{eff-h}(r)=\frac{2m r^3}{r^3+2 m \ell_h^2},
\end{eqnarray}
in which $b$ and $h$ stand for the Bardeen and Hayward metrics, respectively. 
In the extremity $\ell_b,\ell_h\rightarrow\infty$, the effective masses will be equal to $m$. Without loss of generality, by assuming the equatorial plane $(\theta=\pi/2,\dot{\theta}=0)$, the Lagrangian is
\scriptsize{\begin{eqnarray}
\label{eq.10}
2\mathcal{L}&=&-(1-2\frac{M_{eff}(r)}{r})\dot{t}^2+(1-2\frac{M_{eff}(r)}{r})^{-1}\dot{r}^2+r^2\dot{\phi}^2,
\end{eqnarray}}\normalsize
where the symbol "dot" indicates the derivative with respect to the Affine parameter. The generalized momentums are \cite{Zuluaga:2021vjc}
\begin{eqnarray}
\label{eq.011}
   p_{t}&=&\frac{\partial\mathcal{L}}{\partial\dot{t}}=-(1-2\frac{M_{eff}(r)}{r})\dot{t}=-k,\\
\label{eq.012}
   p_{r}&=&\frac{\partial\mathcal{L}}{\partial\dot{r}}=-(1-2\frac{M_{eff}(r)}{r})^{-1}\dot{r},\\
\label{eq.013}
   p_{\phi}&=&\frac{\partial\mathcal{L}}{\partial\dot{\phi}}=r^2\dot{\phi}=h.
\end{eqnarray}
Since the Lagrangian is not an explicit function of $t$ and $\phi$, the momentum correspond $t$ and $\phi$ are invariant. That's why $k$ and $h$, which respectively represent the energy and angular momentum per unit of the particle’s rest mass.\\
Because the Hamiltonian $H=p_t\dot{t}+p_r\dot{r}+p_\phi\dot{\phi}=\mathcal{L}$ is independent of $t$, it is a constant of motion and it goes with
\begin{eqnarray}
\label{eq.014}
2H=-k\dot{t}+(1-2\frac{M_{eff}(r)}{r})^{-1}\dot{r}^2+h\dot{\phi}=-1,
\end{eqnarray}
the second equal is because we are typically deal with material particles \cite{dInverno:1992gxs}. By solving Eqs. (\ref{eq.011}) and (\ref{eq.013}) for $\dot{t}$, $\dot{\phi}$ and substituting in Eq. (\ref{eq.014}), the energy equation is obtained as
\begin{eqnarray}
\label{eq.015}
\frac{1}{2}\dot{r}^2+V_{eff}(r)=\frac{1}{2}(k^2-1),
\end{eqnarray}
where the effective potential per unit mass is given by
\begin{eqnarray}
\label{eq.016}
V_{eff}(r)=-\frac{M_{eff}(r)}{r}+\frac{h^2}{2r^2}-\frac{M_{eff}(r)h^2}{r^3}.
\end{eqnarray}
Here we have another equation which provides $r$ as a function of $\phi$ and describes the orbits of massive particles \cite{M.P. Hobson:2008mnu}.
\begin{eqnarray}
\label{eq.017}
\dot{r}=\frac{dr}{d\tau}=\frac{dr}{d\phi}\frac{d\phi}{d\tau}=\frac{h}{r^2}\frac{dr}{d\phi},
\end{eqnarray}
note that we have used Eq. (\ref{eq.013}).\\
By inserting Eq. (\ref{eq.017}) in Eq. (\ref{eq.015}), we find
\begin{eqnarray}
\label{eq.018}
(\frac{dg}{d\phi})^2+g^2=\frac{k^2-1}{h^2}+\frac{2gM_{eff}(g)}{h^2}+2g^3M_{eff}(g),
\end{eqnarray}
in which the change of variable $r=1/g$ is used. If we substitute the effective mass of Bardeen and Hayward, and then take the derivate from Eq. (\ref{eq.018}) with respect to $\phi$ we obtain the following equations for Bardeen and Hayward metrics, respectively
\begin{eqnarray}
\label{eq.019}
\frac{d^2g}{d\phi^2}+g&=&\frac{m}{(1+\ell_b^2g^2)^{5/2}}[\frac{1}{h^2}(1-2\ell_b^2g^2+3g^2)],
\end{eqnarray}
and
\begin{eqnarray}
\label{eq.020}
\frac{d^2g}{d\phi^2}+g&=&\frac{m}{(1+2\ell_h^2g^3)^2}[\frac{1}{h^2}(1-4\ell_h^2g^3+3g^2)].
\end{eqnarray}
At this point, to determine the basic equations of the time average radial disk structure we need to find the angular momentum $h$, the specific energy $k$ and the angular velocity $\Omega$ of the rotating particles for both black holes. We typically have $\dot{r}=0$ for circular orbits in the equatorial plane, this means that $r$ and  $g = 1/r$ is constant. Therefore, from Eqs. (\ref{eq.019}) and (\ref{eq.020}), we acquire the specific angular momentum
\begin{eqnarray}
\label{eq.021}
h_b&=&mx^2\sqrt{\frac{x^2-2\bar{\ell}_b^2}{(\bar{\ell}_b^2+x^2)^{5/2}-3x^4}},
\end{eqnarray}
and
\begin{eqnarray}
\label{eq.022}
h_h&=&mx^2\sqrt{\frac{x^3-4\bar{\ell}_h^2}{4\bar{\ell}_h^4+4\bar{\ell}_h^2x^3+(x-3)x^5}},
\end{eqnarray}
for simplicity, the above equations are written in practical terms of dimensionless quantities $g=1/mx$ and $\bar{\ell}_b=m\ell_b$ (for Bardeen) and $\bar{\ell}_h=m\ell_h$ (for Hayward). Then, to obtain the specific energy $k$, in Eq. (\ref{eq.015}), we substitute the expression of angular momentum $h$ and set $\dot{r}=0$
\begin{eqnarray}
\label{eq.023}
k_b&=&\frac{(\bar{\ell}_b^2+x^2)^{\frac{3}{2}}-2x^2}{\sqrt{[(\bar{\ell}_b^2+x^2)^{5/2}-3x^4](\bar{\ell}_b^2+x^2)^{\frac{1}{2}}}},
\end{eqnarray}
and
\begin{eqnarray}
\label{eq.024}
k_h&=&\frac{2\bar{\ell}_h^2+(x-2)x^2}{\sqrt{4\bar{\ell}_h^4+4\bar{\ell}_h^2x^3+(x-3)x^5}}.
\end{eqnarray}
By inserting Eqs. (\ref{eq.011}) and (\ref{eq.013}) and using the energy and angular momentum equations for Bardeen and Hayward black holes, the angular velocity is obtained as
\begin{eqnarray}
\label{eq.025}
\Omega_b&=&\frac{\dot{\phi}}{\dot{t}}=\frac{\sqrt{x^2-2\bar{\ell}_b^2}}{m(x^2+\bar{\ell}_b^2)^{5/4}},
\end{eqnarray}
and
\begin{eqnarray}
\label{eq.026}
\Omega_h&=&\frac{\sqrt{x^3-4\bar{\ell}_h^2}}{m(2\bar{\ell}_h^2+x^3)}.
\end{eqnarray}
By inserting Eqs. (\ref{eq.021}) and (\ref{eq.022}) into Eq. (\ref{eq.016}), the effective potential for Bardeen and Hayward black holes can be found as follows respectively 
\begin{eqnarray}
\label{eq.027}
V_{eff-b}&=&-\frac{4\bar{\ell}_b^2x^2\sqrt{\bar{\ell}_b^2+x^2}+x^4(\sqrt{\bar{\ell}_b^2+x^2}-4)}{2[(x^2+\bar{\ell}_b^2)^3-3x^4\sqrt{\bar{\ell}_b^2+x^2}]},
\end{eqnarray}
and
\begin{eqnarray}
\label{eq.028}
V_{eff-h}&=&-\frac{x^2(8\bar{\ell}_h^2+(x-4)x^2)}{2[4\bar{\ell}_h^4+4\bar{\ell}_h^2x^3+(x-3)x^5]}.
\end{eqnarray}
One should notice that in the limit of $\bar{\ell}_b\rightarrow0$ and $\bar{\ell}_h\rightarrow0$, the angular momentum, specific energy, angular velocity and effective potential of the Bardeen and Hayward black holes are reduce to classical expressions.\\
Since circular orbits occur at local minima of the effective potential, the dimensionless radius of the innermost stable circular geodesic orbit $x_{isco}$ can be calculated from \cite{Page:1974he}
\begin{equation}
\label{eq.029}
\frac{d^2V_{eff}}{dx^2}=0
\end{equation}
The derivative of the effective potentials $V_{eff-b}$ and $V_{eff-h}$ are calculated with taking into account that the angular momentum is constant for circular orbits, as presented in Eq. (\ref{eq.016}) \cite{Harko:2009xf,Kovacs:2010xm,Perez:2012bx,Perez:2017spz}
\begin{equation}
\label{eq.030}
\begin{split}
\frac{d^2V_{eff-b}}{dx^2}=&-\frac{2\bar{\ell}_b^4-11\bar{\ell}_b^2x^2+2x^4}{(\bar{\ell}_b^2+x^2)^{7/2}}\\
&+3\bar{h_b^2}(\frac{1}{x^4}+\frac{\bar{\ell}_b^2-4x^2}{(\bar{\ell}_b^2+x^2)^{7/2}}),
\end{split}
\end{equation}
and
\begin{equation}
\label{eq.031}
\begin{split}
\frac{d^2V_{eff-h}}{dx^2}=&-\frac{2(4\bar{\ell}_h^4-14\bar{\ell}_h^2x^3+x^6)}{(2\bar{\ell}_h^2+x^3)^3}\\
&+\bar{h_h^2}(\frac{3}{x^4}+\frac{12x(\bar{\ell}_h^2-x^3)}{(2\bar{\ell}_h^2+x^3)^3},
\end{split}
\end{equation}
for simplicity, $\bar{h}=h/m$ is used. By placing Eqs. (\ref{eq.021}) and (\ref{eq.022}) in Eqs. (\ref{eq.030}) and (\ref{eq.031}), respectively, we obtain
\begin{equation}
\begin{split}
\label{eq.032}
\frac{d^2V_{eff-b}}{dx^2}=&\frac{-2\bar{\ell}_b^4+11\bar{\ell}_b^2x^2-2x^4}{(\bar{\ell}_b^2+x^2)^{7/2}}\\
+&3\frac{(x^2-2\bar{\ell}_b^2)[\bar{\ell}_b^2x^4-4x^6+(\bar{\ell}_b^2+x^2)^{7/2}]}{[(\bar{\ell}_b^2+x^2)^{5/2}-3x^4](\bar{\ell}_b^2+x^2)^{7/2}},
\end{split}
\end{equation}
and
\begin{equation}
\begin{split}
\label{eq.033}
\frac{d^2V_{eff-h}}{dx^2}=&\frac{-2(4\bar{\ell}_h^4-14\bar{\ell}_h^2x^3+x^6)}{(2\bar{\ell}_h^2+x^3)^3}\\
+&3\frac{(x^3-4\bar{\ell}_h^2)[4x^5(\bar{\ell}_h^2-x^3)+(2\bar{\ell}_h^2+x^3)^3]}{(2\bar{\ell}_h^2+x^3)^3(4\bar{\ell}_h^4+4\bar{\ell}_h^2x^3+(x-3)x^5)}.
\end{split}
\end{equation}
Then, using Eqs. (\ref{eq.032}) and (\ref{eq.033}), the value of the innermost stable circular orbit (ISCO) $x_{isco}$ can be computed for the critical values of the free parameters $\bar{\ell}_b=\bar{\ell}_h=\frac{4}{3\sqrt{3}}$ for Bardeen and Hayward black holes, respectively. In addition, $x_{isco}$ can also be calculated from the following relation \cite{Page:1974he}\\
\begin{eqnarray}
\label{eq.034}
\frac{dh}{dx}=\frac{dk}{dx}=0.
\end{eqnarray}
We obtain $x_{isco}=4.82$ for the critical value of the free parameter of the Bardeen black hole $\bar{\ell}_b=\frac{4}{3\sqrt{3}}$, and find that $x_{isco}=5.58$ for the critical value of the free parameter of the Hayward black hole $\bar{\ell}_h=\frac{4}{3\sqrt{3}}$. As we expected the value of the ISCO for the Schwarzschild black hole $\bar{\ell}_b=\bar{\ell}_h=0$ is $x_{isco}=6$.\\
Figure \ref{fig003} shows the effective potential for the constant value $\bar{h}=h/m$ when it is determined at the ISCO for the critical values of the free parameters of Bardeen $\bar{\ell}_b=\frac{4}{3\sqrt{3}}$ (red) and Hayward $\bar{\ell}_h=\frac{4}{3\sqrt{3}}$ (blue). The black dots indicate the locations of the ISCO for each of these cases (Schwartzschild, Bardeen, and Hayward black holes). As it is recognized, for non-zero free parameters $\bar{\ell}_b,\bar{\ell}_h\neq0$ the ISCO (black dots) is shifted towards lesser values of $x$, this displacement is more for Bardeen's black hole compared to Hayward's.\\
In the Fig. \ref{fig004}, we plot the effective potential $V_{eff}$ as a function of $x$ for the same critical values of the free parameters $\bar{\ell}_b=\bar{\ell}_h=\frac{4}{3\sqrt{3}}$ and compared it with the classical solution. One can infer from this figure and from the values obtained for $x_{isco}$ that for regular black holes the radius of the ISCO and the values of the minimum effective potential shift to lesser values. As it is clear from the figure, the displacement of the Bardeen black hole is more than Hayward black hole.\\
\begin{figure}
\centering
\includegraphics[width=1\linewidth]{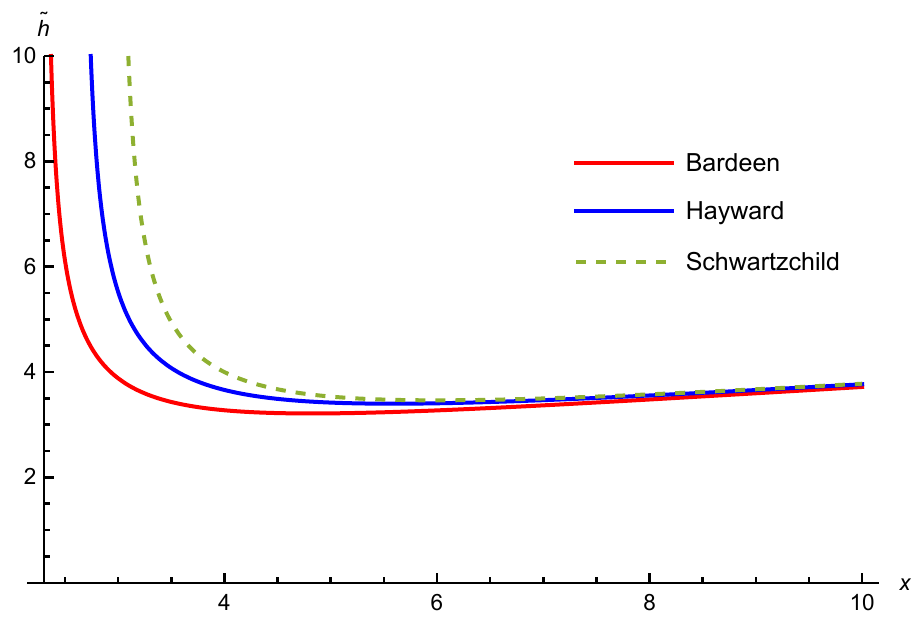}
\caption{Angular momentum $\bar{h}=h/m$ as a function of $x$ for equal and critical values of free parameters $\bar{\ell}_b=\bar{\ell}_h=\frac{4}{3\sqrt{3}}$. The red curve shows the Bardeen black hole, the blue curve shows the Hayward black hole, and the dashed curve shows the classical $\bar{\ell}_b=\bar{\ell}_h=0$ state.}
\label{fig005}
\end{figure}
The angular momentum curve $\bar{h}$ as a function of $x$ for the critical values of the free parameters $\bar{\ell}_b=\bar{\ell}_h=\frac{4}{3\sqrt{3}}$ is drawn in the Fig. \ref{fig005} and compared with the classical solution. From examining the angular momentum curve and the values obtained for $x_{isco}$, one can see that in addition to the fact that the radius of the ISCO and the minimum effective potential have been shifted to a lesser value of $x$, the angular momentum has also been slightly shifted to lesser values in the vicinity of $x_{isco}$. It is clear from the figure, the shift value is higher for the Bardeen black hole compared to the Hayward black hole.\\
According to Fig. \ref{fig006}, we see that the same behavior as angular momentum is repeated in the curve of specific energy $k$ in term of $x$. Figure 6 which is plotted for the critical values of the free parameters l $\bar{\ell}_b=\bar{\ell}_h=\frac{4}{3\sqrt{3}}$ and compared with the classical solution, shows that the specific energy k of regular black holes is slightly shifted towards lesser values compared to the classical case. The magnitude of this shift is more considerable for the Bardeen black hole compared to the Hayward black hole, similar to the angular momentum.\\
\begin{figure}
\centering
\includegraphics[width=1\linewidth]{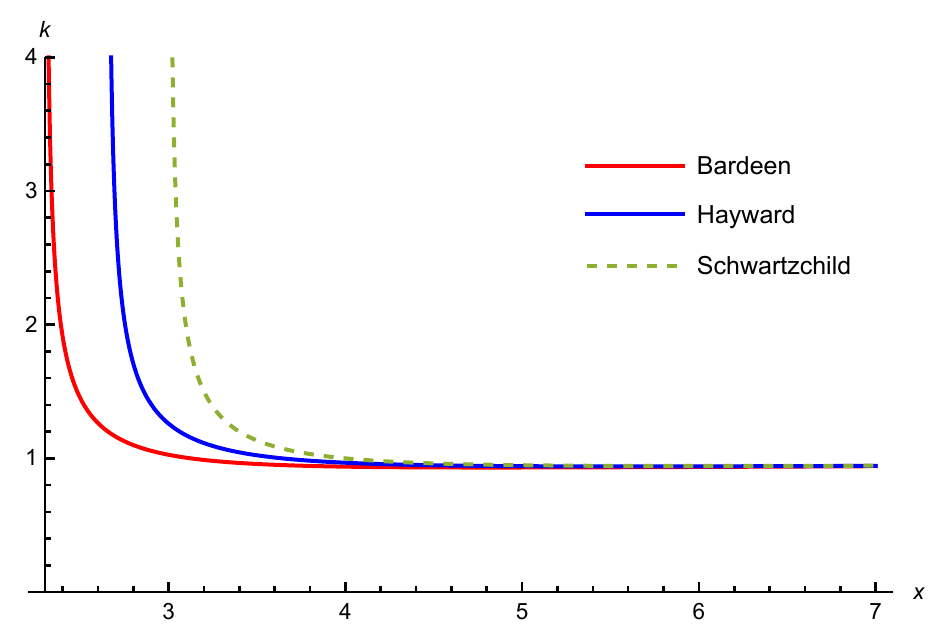}
\caption{Specific energy $k$ as a function of $x$ for equal and critical values of free parameters $\bar{\ell}_b=\bar{\ell}_h=\frac{4}{3\sqrt{3}}$. The red curve shows the Bardeen black hole, the blue curve shows the Hayward black hole, and the dashed curve shows the Schwarzschild black hole $\bar{\ell}_b=\bar{\ell}_h=0$.}
\label{fig006}
\end{figure}
Equation (\ref{eq.016}) together with Eqs. (\ref{eq.021}) and (\ref{eq.022}) show that the combination of the two effects: 1)  the mass of the regular black hole is lesser than the mass of the Schwarzschild black hole $M_{eff}<m$ and 2) the values of angular momentum $h$ decrease for non-zero values of free parameters $\bar{\ell}_b=\bar{\ell}_h\neq0$) deepens the potential well and shifts the ISCO $x_{isco}$ to lesser values.\\
Figure \ref{fig007} shows the effective mass curve of two black holes Bardeen (red line) and Hayward (blue line), as a function of $x$ and compares it with the classical case (horizontal green dashed line). As it can be seen from it, for a certain value of the free parameters $\bar{\ell}_b$ and $\bar{\ell}_h$, the particle falling into the black hole feels the mass $M_{eff}$, which decreases as the $x$ values decrease. This is while the mass of the Schwarzschild black hole $M_{eff}$ is constant and does not change with decreasing $x$. Thus, although the angular momentum is also reduced, the particle feels a weaker gravitational pull as it spirals into the black hole, allowing it to remain in a stable circular orbit with a smaller $x_{isco}$. It is also clear from the curve obtained for the two black holes that compared to the effective mass of the Hayward black hole, the effective mass of the Bardeen black hole decreases at a larger $x$ but with a lower slope.\\
\begin{figure}
\centering
\includegraphics[width=1\linewidth]{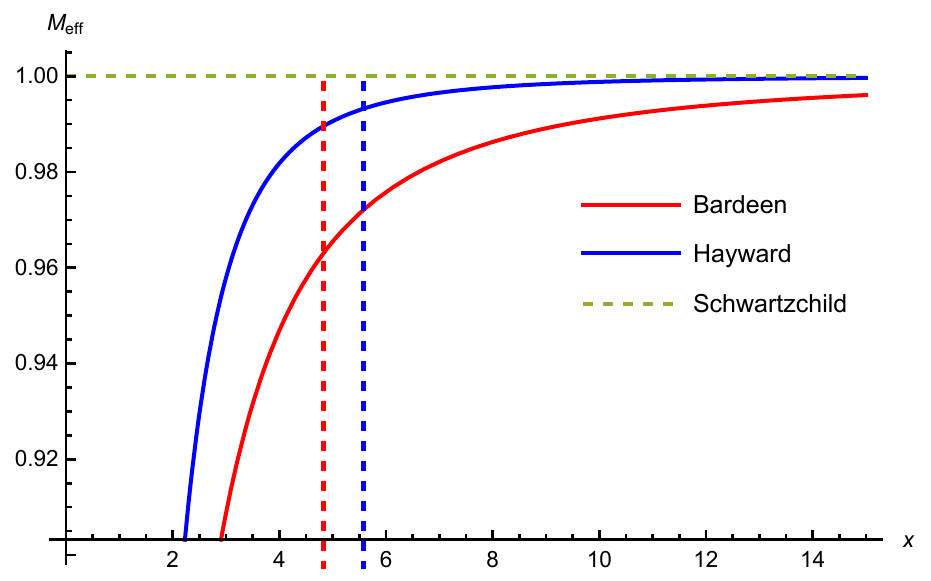}
\caption{Effective mass curves as a function of $x$ for Bardeen (red) and Hayward (blue) black holes. The vertical dashed lines show the ISCO for the Bardeen and Hayward holes for the limit values of the free parameters $\bar{\ell}_b=\bar{\ell}_h=\frac{4}{3\sqrt{3}}$, respectively. The horizontal dashed line corresponds to the classical case $\bar{\ell}_b=\bar{\ell}_h=0$.}
\label{fig007}
\end{figure}

\section{Relativistic thin accretion disk}

\begin{figure}
\centering
\includegraphics[width=1\linewidth]{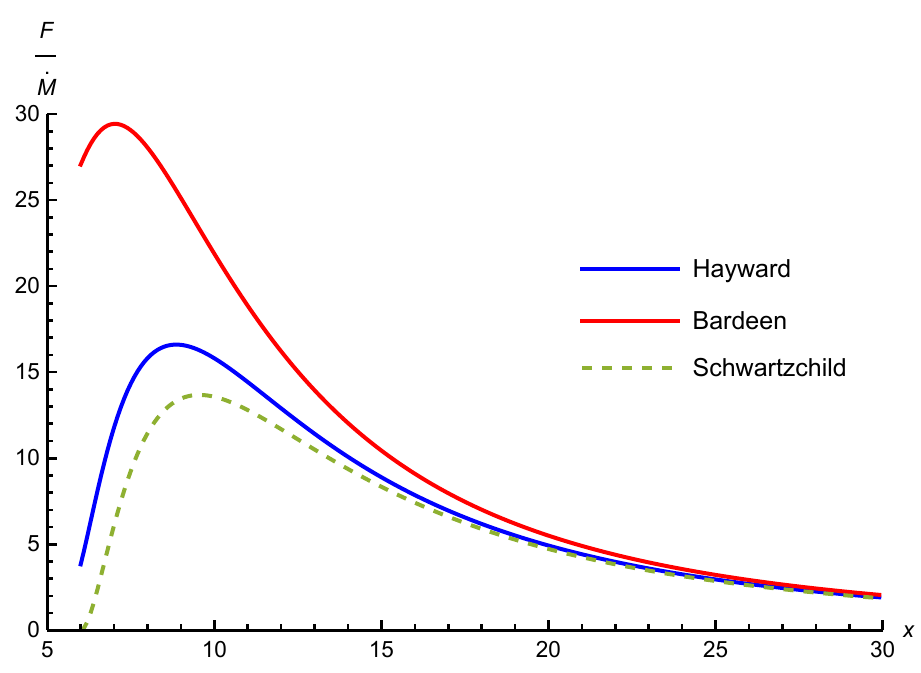}
\caption{Radiant energy flux per unit disk accretion rate from a thin accretion disk around the Bardeen black holes $\bar{\ell}_b=\frac{4}{3\sqrt{3}}$ (red) and Hayward $\bar{\ell}_h=\frac{4}{3\sqrt{3}}$ (blue) along with the disk radiative energy flux around the Schwarzschild classical black hole $\bar{\ell}_b=\bar{\ell}_h=0$ (line green curve) is depicted.}
\label{fig008}
\end{figure}

\begin{figure}
\centering
\includegraphics[width=1\linewidth]{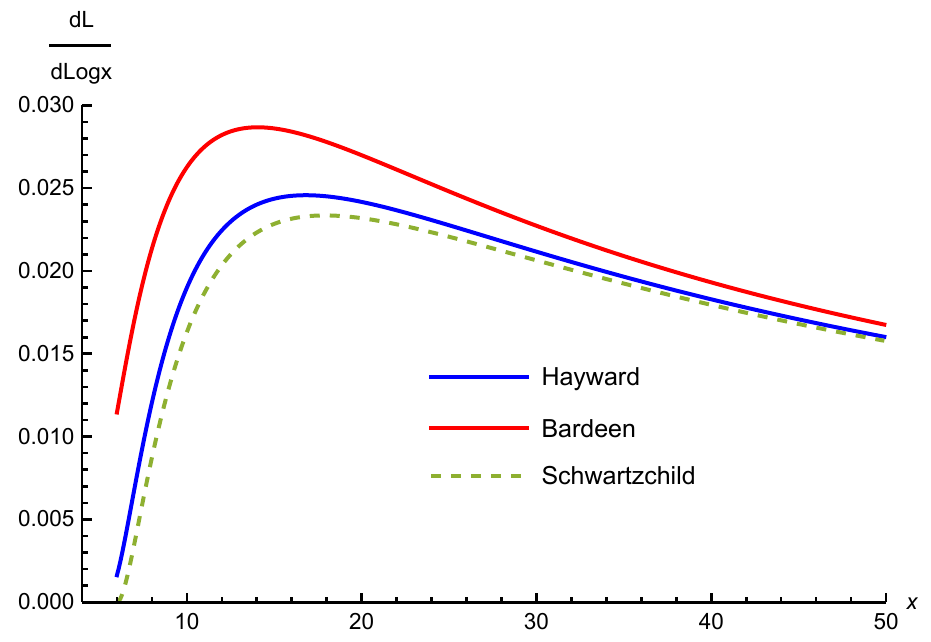}
\caption{The luminosity derivative at infinity in the accretion rate unit of the thin disk around the Bardeen $\bar{\ell}_b=\frac{4}{3\sqrt{3}}$ (red) and Hayward $\bar{\ell}_h=\frac{4}{3\sqrt{3}}$ (blue) black holes along with the luminosity derivative at infinity in the accretion rate unit around the classical black hole $\bar{\ell}_b=\bar{\ell}_h=0$ (green dashed line).}
\label{fig009}
\end{figure}

\begin{figure}
\centering
\includegraphics[width=1\linewidth]{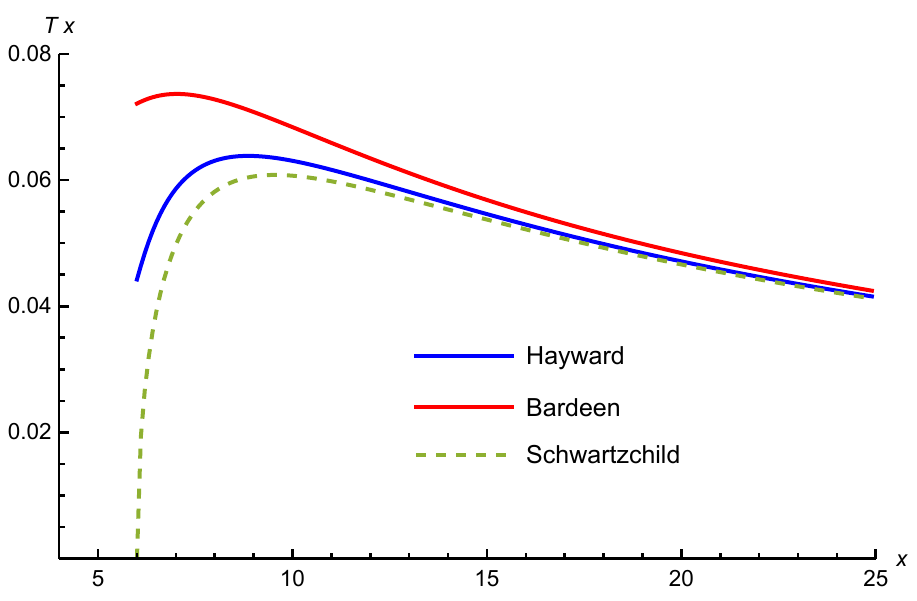}
\caption{The temperature in the accretion rate unit of the disk around Bardeen $\bar{\ell}_b=\frac{4}{3\sqrt{3}}$ (red) and Hayward $\bar{\ell}_h=\frac{4}{3\sqrt{3}}$ (blue) black holes along with the temperature of the disk in the accretion rate unit around the Schwarzschild $\bar{\ell}_b=\bar{\ell}_h=0$ black hole (dashed green line).}
\label{fig010}
\end{figure}

In the most conventional non-relativistic model of an accretion disk around a compact central body, it is assumed that matter spiraling towards the center of the disk loses angular momentum due to turbulent viscosity that is transported outward through the disk. As the gas moves toward the center, it loses gravitational energy and heats up, emitting heat energy \cite{Shakura:1973boa}. For the first time, the general relativistic behavior of accretion disks around black holes was studied in \cite{Novikov:1973kta,Page:1974he}. The accretion disk is thin, meaning that the half-width of the accretion disk $H$ is extremely small, so the disk radius is considerably larger than its width $H/R\ll1$ (where $R$ is the characteristic disk radius). In this case, the heat produced by dynamic tension and friction is effectively emitted from the disk surface through radiation. The quantities describing the thermal properties of the disk are averaged over the azimuthal angle $\phi= 2\pi$, the height $H$, and the time scale $\Delta t$ (the time it takes for the gas to flow inwards over a distance of $2H$). According to the specified assumptions, the radial structure of the time average of the disk is obtained from the conservation laws of rest mass, energy, and angular momentum. By integrating from the equation of conservation of mass, we find the constancy of the rate of mass increase as \cite{Zuluaga:2021vjc}

\begin{eqnarray}
\label{eq.035}
\dot{M}=-2\pi r \Sigma (r)u^r=constant,
\end{eqnarray}
In the above relation, $\Sigma(r)$ represents the surface density of the disk and $u^r$ denotes the radial velocity. By combining the laws of conservation of energy and angular momentum, the derivative of luminosity at infinity $L_{\infty}$ is found as \cite{Page:1974he,Joshi:2013dva}
\begin{eqnarray}
\label{eq.036}
\frac{d\mathcal{L}_{\infty}}{dlnr}=4\pi r\sqrt{-g}k\mathcal{F}(r),
\end{eqnarray}
which in the local framework of the accretion fluid, the radiant energy flux $\mathcal{F}$ emitted from the upper surface of the disk, in terms of angular momentum $h$, specific energy $k$, and angular velocity $\Omega$ is given by
\scriptsize{\begin{eqnarray}
\label{eq.037}
\mathcal{F}(r)=-\frac{\dot{M}}{4\pi \sqrt{-g}}\frac{1}{(k-\Omega h)^2}\frac{d\Omega}{dr}\times\int_{r_{isco}}^r(k-\Omega h)\frac{dh}{dr}dr,
\end{eqnarray}}\normalsize
where $\sqrt{-g}=r$ applies to both regular black hole metrics and classical Schwarzschild spacetime. The numerical integration of Eq. (\ref{eq.037}) will be easier by implementing an integration by parts and using the relation $dk/dr=\Omega(dh/dr)$ \cite{Page:1974he}
\scriptsize{\begin{eqnarray}
\label{eq.038}
\int_{r_{isco}}^r(k-\Omega h)\frac{dh}{dr}dr=kh-k_{isco}h_{isco}-2\int_{r_{isco}}^rh\frac{dk}{dr}dr.
\end{eqnarray}}\normalsize
Since the disk is assumed to be in thermodynamic equilibrium, the radiation emitted from the surface of the accretion disk can be considered as blackbody radiation with a temperature given by \cite{Zuluaga:2021vjc}
\begin{eqnarray}
\label{eq.039}
T(r)=\sigma^{-\frac{1}{4}}\mathcal{F}(r)^{\frac{1}{4}},
\end{eqnarray}
where $\sigma$ is the Stefan-Boltzmann constant.\\
The black hole mass $m$ and the rate of mass increase $\dot{M}$ are considered observational constants \cite{Benedetti:2009rx,Cai:2010zh,Falls:2017lst,Zhang:2018xzj}, and to be able to compare the differences easily, the mass is considered equal to the unit $m = 1$, and the calculation of the thermal properties of the disk is performed in the unit of mass accretion rate \cite{Itin:2008mnu}.\\ 
In Figs. \ref{fig008}, \ref{fig009} and \ref{fig010}, we draw the radial curve of the time average energy flux, the derivative curve of the luminosity, and the temperature curve of the accretion disk in the accretion rate unit for the critical value of the free parameters of the Bardeen and Hayward black holes $\bar{\ell}_b=\bar{\ell}_h=\frac{4}{3\sqrt{3}}$ and the classical case $\bar{\ell}_b=\bar{\ell}_h=0$.\\
Using Fig. \ref{fig008}, we compare the radiation energy flux from the accretion disk of regular black holes with the radiation energy flux from the accretion disk of a classical black hole. It is clear more energy is emitted from the accretion disk around regular black holes compared to the Schwarzschild black hole. This is because the inner edge of regular black holes is shifted towards lesser values of $x$ compared to the classical case. Also, as expected, the radiative energy flux of the Bardeen black hole is higher than the radiative energy flux of the Hayward black hole. The radiant energy flux of Berdeen black hole is 115\%, and Hayward black hole is 21\% more than Schwarzschild black hole.\\
Fig. \ref{fig009} shows that the derivative of the luminosity of the regular black hole's accretion disk also changes compared to the Schwarzschild black hole. In this figure, the luminosity derivative curve for both Bardeen and Hayward black holes is plotted for the critical value of the free parameters $\bar{\ell}_b=\bar{\ell}_h=4/3\sqrt{3}$ and compared with the luminosity derivative of the classical black hole. Similar to the radiative energy flux, here, the derivative of the Bardeen black hole's luminosity is higher than the Hayward black hole and is closer to smaller values of $x$. Compared to the derivative of the Schwarzschild black hole's luminosity derivative of the accretion disk of the Bardeen and Hayward black holes is 22 and 5\% higher, respectively.\\
Figure \ref{fig010} is related to the temperature of accretion disks in the unit of disk accretion rate. In this figure, the temperature curve of the accretion disk around two regular black holes Bardeen $\bar{\ell}_b=\frac{4}{3\sqrt{3}}$ (red curve) and Hayward $\bar{\ell}_h=\frac{4}{3\sqrt{3}}$ (blue curve) is compared with the temperature of the accretion disk around the classical black hole $\bar{\ell}_b=\bar{\ell}_h=0$ (dashed green line). We can see from the figure, due to the displacement of the inner edge of the accretion disk in regular black holes towards lesser values of $x$, the temperature of Bardeen and Hayward black holes is higher compared to Schwarzschild black hole. Also, by comparing two regular black holes, Bardeen and Hayward, it is clear that the accretion disk temperature is higher in Bardeen black hole compared to Hayward black hole (21\% and 5\% more than the classic mode, respectively). This issue is related to the lessness of the ISCO of the Bardeen black hole ($x_{isco}=4.82$) compared to the Hayward black hole ($x_{isco}=5.58$).\\
\section{Conversion efficiency of mass to radiation}
Under the condition that all photons emitted from the surface of the disk can escape to infinity, we obtain the accretion mass-to-radiation conversion efficiency $\epsilon$ using the energy dissipation by the test particle moving from infinity towards the inner boundary of the disk. Therefore, for $r\rightarrow\infty$, $k_\infty\approx1$, we have
\begin{eqnarray}
\label{eq.040}
\epsilon=\frac{k_\infty-k_{isco}}{k_\infty}\approx1-k_{isco},
\end{eqnarray}
For the critical values of the free parameters $\bar{\ell}_b=\bar{\ell}_h=\frac{4}{3\sqrt{3}}$, we calculate the values of the specific energy in the innermost stable circular orbit $k_{isco}$ and the values of the conversion efficiency of the accretionary mass into the corresponding radiation $\epsilon$. As one expects, and it is comprehensible from Table \ref{table:1}, the value of conversion efficiency of accretion mass to radiation $\epsilon$ increases for nonzero values of free parameters, in other words, for regular black holes compared to Schwarzschild black hole. This means that, in general, regular black holes are a more efficient engine for converting cumulative mass into radiation than Schwarzschild black holes.\\
\begin{table}[h!]
\centering
\begin{tabular}{ l | c | r }
  \hline			
  Free Parameter & $k_{isco}$ & $\epsilon(\%)$ \\
  \hline
  $\bar{\ell}_b=4/3\sqrt{3}$ & 0.9314 & 6.8568 \\
  $\bar{\ell}_h=4/3\sqrt{3}$ & 0.9397 & 6.0248 \\
  $\bar{\ell}_b=\bar{\ell}_h=0$ & 0.9428 & 5.7191 \\
  \hline
\end{tabular}
  \caption{Specific energy in the innermost stable circular orbit $k_{isco}$ and mass-to-radiation conversion efficiency $\epsilon$ for the critical value of the free parameters of Bardeen and Hayward black holes and comparison with the classical Schwarzschild mode.}
\label{table:1}
\end{table}
In addition, comparing two regular black holes, we conclude that the conversion efficiency of accretion mass into $\epsilon$ radiation is higher for the Bardeen black hole than for the Hayward black hole.\\
The increase in the maximum of the radiative energy flux, temperature, and luminosity derivative, as well as the increase in the conversion efficiency of accretionary mass to radiation, are directly related to the deeper potential well of regular black holes and the displacement of the ISCO to lesser values.\\

\section{Black hole with non-zero spin}

\begin{figure}
\centering
\includegraphics[width=1\linewidth]{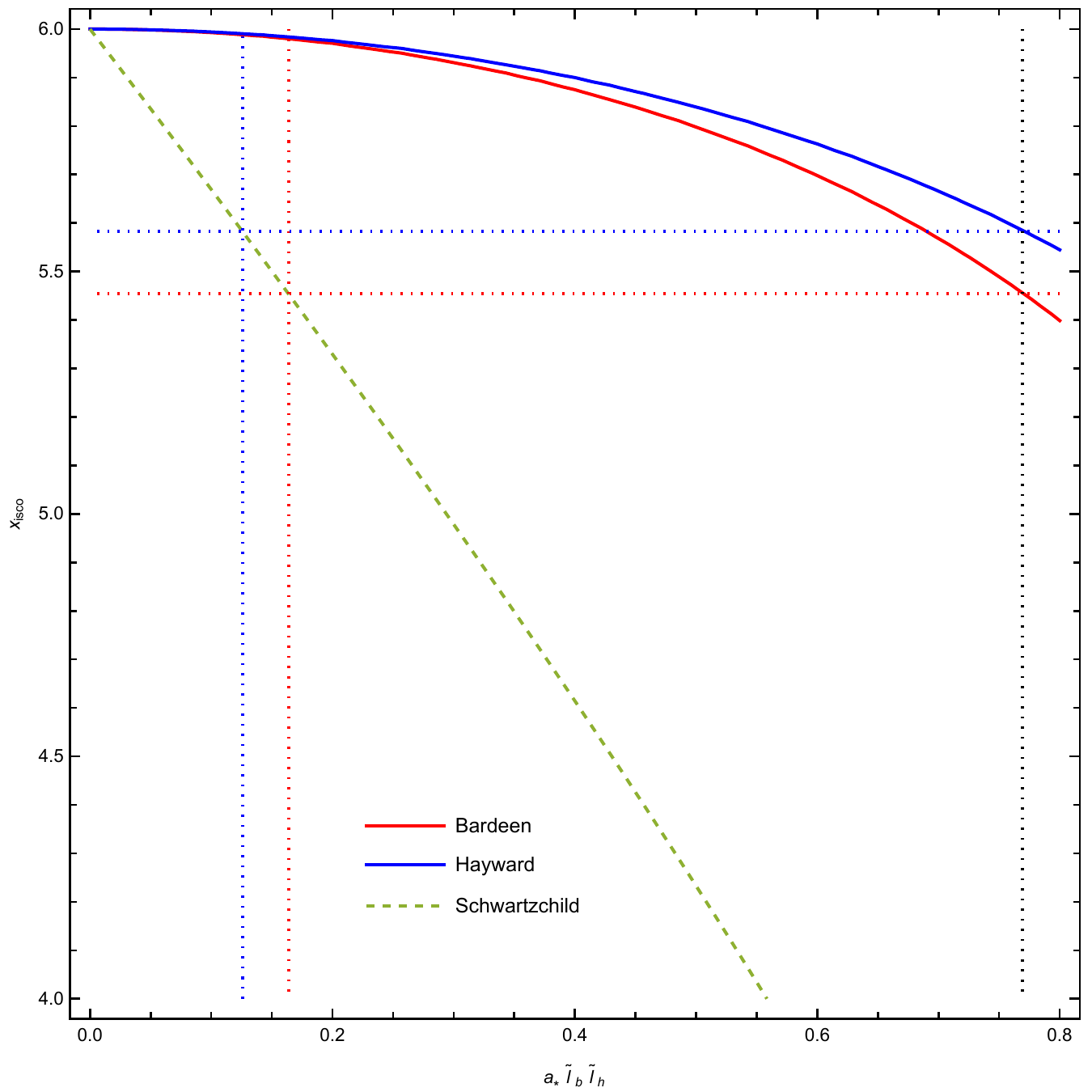}
\caption{Dependence of the radius of the innermost stable circular orbit $x_{isco}$ on the spin parameter $a_*$ for the Kerr black hole (green dashed line) and the free parameter b for the Bardeen black hole (red curve) and the free parameter h for the Hayward black hole (blue curve).}
\label{fig011}
\end{figure}

We know that the smallness of the ISCO in rotating black holes can be related to their non-zero spin \cite{Zuluaga:2021vjc}. Based on this issue, the possible astrophysical application of the results can be investigated. For this practical purpose, it should be seen whether the free parameters $\bar{\ell}_b$ and $\bar{\ell}_h$ can behave properly similarly to the non-zero spin of Kerr black holes? That is, whether the spin parameter $a^*=a/M$ and the dimensionless free parameters $\bar{\ell}_b$ and $\bar{\ell}_h$ can create the same radius of the ISCO or not? We know that the radius of the ISCO for circular orbits around the Kerr black hole is given as \cite{Bardeen:1972fi}
\begin{eqnarray}
\label{eq.041}
r_{isco}=3+Z_2-\sqrt{(3-Z_1)(3+Z_1+2Z_2)},
\end{eqnarray}
in which we have
\begin{eqnarray}
\label{eq.042-043}
Z_1&=&1+\sqrt[3]{1-a^2}(\sqrt[3]{1-a}+\sqrt[3]{1+a}),\\
Z_2&=&\sqrt{3a^2-Z_1^2}.
\end{eqnarray}
Also, the ISCOs of the thin accretion disk around Bardeen and Hayward black holes are obtained from the roots of Eqs. (\ref{eq.032}) and (\ref{eq.033}), respectively
\begin{equation}
\begin{split}
\label{eq.044}
(-2\bar{\ell}_b^4+11\bar{\ell}_b^2x^2-2x^4)(-3x^4+(\bar{\ell}_b^2+x^2)^{5/2})\\
+3(x^2-2\bar{\ell}_b^2)(\bar{\ell}_b^2x^4-4x^6+(\bar{\ell}_b^2+x^2)^{7/2})=0,
\end{split}
\end{equation}
and
\begin{equation}
\begin{split}
\label{eq.045}
-2(4\bar{\ell}_h^4-14\bar{\ell}_h^2x^3+x^6)(4\bar{\ell}_h^4 +4\bar{\ell}_h^2x^3+(x-3)x^5)\\
+3(2\bar{\ell}_h^2+x^3)^3)+(x^3-4\bar{\ell}_h^2)(12x^5(\bar{\ell}_h^2-x^3)=0.
\end{split}
\end{equation}
We have plotted the relation between the spin of the Kerr black hole $a_*$ and the dimensionless parameters of the regular black holes $\bar{\ell}_b$ and $\bar{\ell}_h$ in the Fig. \ref{fig011}. The two red and blue curves indicate the dependence of the radius of the innermost stable circular orbit $x_{isco}$ on the free parameters $\bar{\ell}_b$ and $\bar{\ell}_h$ of Bardeen and Hayward black holes, respectively. The green dashed line corresponds to the spin of the Kerr black hole. In this diagram, the vertical black dotted line shows the critical value of the free parameters $\bar{\ell}_b$ and $\bar{\ell}_h$. The intersection of the red (Bardeen) and blue (Hayward) vertical dashes with the horizontal axis allows us to conclude that the free parameters $\bar{\ell}_b$ and $\bar{\ell}_h$ can mimic the spin of the Kerr black hole up to a maximum of $a_{*b}=0.164$ and $a_{*h}=0.126$, respectively. As it is clear from the figure, these maximum values are got for the critical values of the free parameters $\bar{\ell}_b$ and $\bar{\ell}_h$. Therefore, one can conclude that Bardeen and Hayward regular black holes can imitate Kerr black holes to a certain extent. Compared to the Hayward black hole, the Bardeen black hole can mimic the behavior of the Kerr black hole up to a greater spin.\\
As an interesting example of the astrophysical application of the performed calculations, we can mention the black hole LMC X-3. This black hole, located outside the Milky Way and in an X-ray binary system, has a stellar mass of $M=6.98\pm 0.56M_\odot$ \cite{Steiner:2014ufa}. The calculated spin parameter for this black hole is equal to $a_*=0.21_{+0.18}^{-0.22}$ with 90\% CL \cite{Steiner:2014zha}. So, assuming that the accretion disk around the LMC X-3 black hole is a relativistic thin disk and considering that the value obtained for $a_*^{AS}$ is within the uncertainty range of the value obtained from the direct observations, we can fix the dimensionless free parameters on the critical values $\bar{\ell}_b=\bar{\ell}_h=\frac{4}{3\sqrt{3}}$. However, it should be noted that observational data show that the relativistic thin disk model is an accurate model at low luminosities, but not at high luminosities. For an accretion disk at high luminosities, the slim accretion disk model provides a better description \cite{Steiner:2010kd,Straub:2011ii}.\\
The obtained results show regular black holes do not affect only the interior of the black hole horizon and its vicinity. Rather, these effects remain beyond the boundaries of the black hole horizon and even cause changes in the thermal properties of the accretion disk around the black hole. Therefore, studying regular black holes in more realistic environments such as accretion disks around them can be useful for a more proper understanding of black holes and general relativity.

\section{Conclusion}

In this paper, we studied the thermal properties of relativistic thin accretion disks around regular black holes. For this purpose, we calculated the time-averaged energy flux corrections, the derivative of the luminosity at infinity, the disk temperature, and the conversion efficiency of accretion mass into radiation. We found that compared to the predictions of general relativity, the non-zero free parameters $\bar{\ell}_b$ and $\bar{\ell}_h$ cause the radius of the ISCO of the disk to shift to lesser values. It was also found that both Bardeen and Hayward black holes have only one event horizon per critical value of free parameters ($\bar{\ell}_b=\bar{\ell}_h=\frac{4}{3\sqrt{3}}$). While for free parameters greater than the critical value, none of the regular black holes have a horizon ($\bar{\ell}_b=\bar{\ell}_h>\frac{4}{3\sqrt{3}}$). On the other hand, for values lesser than the critical value of the free parameters, both Bardeen and Hayward black holes have two horizons ($\bar{\ell}_b=\bar{\ell}_h<\frac{4}{3\sqrt{3}}$), one (internal) Cauchy horizon and one (external) event horizon. Meanwhile, keeping the free parameters constant in critical value and changing the mass has a different effect on the horizon of the black hole. In fact, changing the mass changes the location of the event horizon, not its number. In this way, as the mass decreases, the horizon of the black hole moves towards lesser values of $r$. However, in any particular case, the horizon corresponding to the Bardeen black hole is smaller than the horizon corresponding to the Hayward black hole.\\
We found the increase in energy radiated from the surface of the disk around the black hole is another direct result of the non-zero free parameters $\bar{\ell}_b$ and $\bar{\ell}_h$. We also detect an increase in the temperature of the accretion disk around regular black holes for non-zero values of the free parameters $\bar{\ell}_b$ and $\bar{\ell}_h$. In addition, the non-zero free parameters has caused an increase in the derivative of the luminosity of the accretion disk around the Bardeen and Hayward black holes. The increase in all quantities is related to the smaller radius of the ISCO of the accretion disk around the black hole. By looking carefully at the radiant energy, temperature and luminosity derivative profiles, we notice that for all the quantities, in addition to the increase of the peak, the peak of the radial profiles has also shifted towards smaller values of $x$. In all the above cases, the displacement of the Bardeen black hole is greater than the displacement of the Hayward black hole.\\ 
On top of that, we saw an increase in the conversion efficiency of accretion mass to radiation for non-zero values of free parameters compared to the classical case. That is, regular black holes are more efficient for converting mass into radiation compared to Schwarzschild black holes.\\
Another result of this paper is related to the spin of black holes. In this way, the Bardeen and Hayward regular black holes can imitate the spin parameter of the Kerr black hole up to the values of $a_{*b}=0.164$ and $a_{*h}=0.126$, respectively. These values are obtained when the dimensionless free parameters take their critical values. Finally, considering that the value obtained for the spin parameter of both black holes is within the uncertainty range of the spin of the LMC X-3 black hole, we applied our findings to this black hole. So maybe the LMC X-3 black hole can be introduced as a black hole without intrinsic singularity with $\bar{\ell}=\frac{4}{3\sqrt{3}}$. However, since we performed the calculations under the assumption that the accretion is from the relativistic thin disk model, this claim should be made with caution.\\
Finally, we remind that the effect of the regularity of a black hole is not limited to the horizon of black holes or its vicinity and appear at distances greater than the radius of the ISCO. This means that studying the accretion disks around various types of black holes and comparing them with observations can be useful in a better understanding of black holes.

%
%

\end{document}